\documentstyle[11pt,newpasp,twoside,epsf]{article}

\def\sun{\hbox{$\odot$~}}
\def\arcmin{\hbox{$^\prime$}}
\def\arcsec{\ifmmode^{\prime\prime}\;\else$^{\prime\prime}\;$\fi}

\def\hawaii{Hawai$'$i~}

\def\rx{RXJ\thinspace 1716.6$+$6708}

\begin{document}

\title{RXJ1716.6+6708: a protocluster at z$=$0.81?}
\author{Isabella M. Gioia}
\affil{Istituto di radioastronomia del CNR, 40129 Bologna, Italy}
\affil{Institute for Astronomy, Honolulu, HI, 96822, USA}
\begin{abstract}
\end{abstract}
At z$=$0.809,  \rx ~is the second most distant X-ray selected 
cluster so far published and the only one with a large number of
spectroscopically determined cluster member velocities. 
The optical morphology of \rx ~resembles an inverted S-shape
filament with the X-rays coming  from the midpoint of the filament. 
The ROSAT HRI contours have an elongated shape that roughly coincide 
with the weak lensing contours. ASCA measures a low temperature,
kT=5.7 keV. Keck-II LRIS spectra indicate
a very high velocity dispersion, $\sigma_{los}=1522$  km s$^{-1}$.
While the  temperature is commensurate with its X-ray luminosity,
its velocity dispersion is much higher than expected 
from the $\sigma-T_{X}$ relationship  of present-day clusters with 
comparable X-ray luminosity.  \rx ~could be an example of a
protocluster, where matter is flowing along filaments and the X-ray
flux is maximum at the impact point of the colliding streams of matter.


\section{Introduction}

Clusters of galaxies at redshifts nearing one are of special
importance  since they may be caught at the epoch of
formation. In hierarchical theories of structure formation,
clusters of galaxies form from the high peaks in the original density 
field, thus they provide crucial constraints on the shape, amplitude, and
temporal evolution of the primordial mass fluctuation spectrum.
Despite their importance, the statistics for the abundance of high-z 
($\sim$0.8 and beyond) clusters are poor since they are so difficult 
to locate. One of the cleanest ways to avoid  sample contamination
is  the selection of high-z clusters  by means
of their X-ray emission. X-ray surveys are sensitive enough
to detect distant clusters. Examples include MS1137$+$66,
at z$=$0.78 and MS1054$-$03 at z$=$0.83 in the Medium Survey
(Gioia et al. 1990; Gioia and Luppino, 1994; see also the detailed 
study on MS1054$-$03 by Donahue et al. 1996).
Other X-ray surveys being conducted with ROSAT archive data
are also  finding distant clusters, for example the Wide Angle
ROSAT Pointed Survey cluster at 0.83 (RXJ0152.7-1357, Ebeling et al. 
1999) and the clusters discovered in the ROSAT Deep Cluster Survey 
(Rosati et al. 1998; Rosati et al. 1999). It is worth noticing that
these high-z massive clusters are filamentary in optical with the
X-rays following the elongation of the optical galaxies in most cases.
Velocity dispersions and temperatures, when available, show high values. 
Are we starting to observe the formation epoch of massive clusters?

\section{The NEP Survey}
Our group at the Institute for Astronomy in \hawaii
has been involved for several years in the optical
identification of all the sources found in 
the North Ecliptic Pole (NEP) region  of the  ROSAT
All-Sky Survey (RASS, Tr\"umper et al. 1991; Voges et al. 1999). 
The NEP region is the deepest area of the RASS where the 
ROSAT satellite scan circles overlap and where the effective 
exposure time exceeds 35ks. The 9-year long identification program 
has been finally completed. A distant cluster, named \rx, at 
z$=$0.81 was detected in the NEP with only 33 net photons (Henry et al. 1997;
Gioia et al. 1999). This weak detection revealed a very interesting 
object.  Keck-II spectroscopy and X-ray follow-up observations with 
the ROSAT High Resolution Imager (HRI) and with the ASCA satellite were 
performed and are presented here and, in more detail, in Gioia et al. (1999).
Throughout this paper, we assume H$_{0}=50$
km s$^{-1}$ Mpc$^{-1}$, and q$_{0}=0.5$.

\section{Observations}

Fig. 1 presents the optical image of \rx. The cluster morphology is 
elongated in the NE-SW direction. Optical spectroscopy with the Low 
Resolution and Imaging Spectrograph (Oke et al. 1995) on the Keck-II 
telescope provided redshifts for 37 cluster members. An average redshift
z$=$0.8090$\pm$0.0051 and velocity dispersion along the line of sight
$\sigma_{los}=$1522$^{+215}_{-150}$  km s$^{-1}$ were obtained.
From ASCA data a best-fit cluster rest frame  temperature of 
$kT=5.7^{+1.37}_{-0.58}$ keV was measured for the cluster gas and  
a value A$=0.43^{+0.25}_{-0.21}$ for the metallicity. The HRI reveals 
that \rx ~is morphologically complex at X-ray wavelengths too (see Fig. 2). 
The emission is clearly extended over a scale of roughly 1 arcmin 
($\approx 500$ h$^{-1}_{50}$ kpc), and its shape is indicative of 
the non-regularity of this cluster. A total flux 
f$_{0.5-2keV}=$(1.66$\pm$0.09)$\times10^{-13}$ erg cm$^{-2}$ s$^{-1}$, and
a luminosity L$_{0.5-2keV}=$(4.57$\pm$0.24)$\times10^{44}$
h$_{50}^{-2}$ erg s$^{-1}$ are obtained.  From the HRI surface 
brightness profile,  assuming a constant temperature, kT=5.7 keV, 
as measured by ASCA, the three-dimensional density distribution of 
the gas from the two-dimensional image was derived.
Our estimates for the gas and gravitational mass within  
$\sim$1h$_{50}^{-1}$ Mpc using HRI data are (8.9$\pm2.1)\times10^{13}$ 
h$_{50}^{-5/2} M_{\sun}$ and (2.8$\pm0.3)\times10^{14}$ h$_{50}^{-1} M_{\sun}$.
The total mass from X-rays is lower, even within the uncertainties, than
the mass derived from weak lensing by Clowe et al. (1998),
M$_{wl}=$(5.2$\pm$1.8)$\times10^{14}$ h$_{50}^{-1} M_{\sun}$.
We note however that the Clowe et al. (1998)  mass determination 
includes the second clump of matter associated with the NE group of
galaxies (see Fig. 2). This second clump is clearly separated in the 
Clowe et al. mass distribution map from the central cluster mass, and
is not detected in the HRI.

\section{Discussion}

There are not many X-ray selected clusters at high redshift. The other
published examples are the two clusters at t z$\sim$0.8 from the EMSS.
Similarly to  MS1054$-$03, \rx~ has  a high velocity dispersion
which can be interpreted as a signature of non-virialization.
\rx ~has an extended X-ray emission of about 1\arcmin,  centered 
on the cD galaxy and with an elongation in the same direction as
the optical galaxies. The temperature of \rx ~is commensurate
with the predictions from its X-ray luminosity from the
L$_{X}-$T$_{X}$ relation by  David et al. (1993) and 
Arnaud and Evrard (1999). The  measured velocity dispersion however
is much higher than expected from its temperature. Using the derived
value from the HRI of  $\beta$ = 0.42 and  the measured ASCA temperature 
in the equation  $\beta = { \mu m_p \sigma_v^{2} \over k T_{gas} }$
a velocity dispersion of about 600 km s$^{-1}$ is obtained, much
lower than measured. Conversely, using the $\sigma-T_X$ relationship
(Girardi et al. 1996) a temperature of  of 11.5$^{+3.0}_{-1.6}$ keV would
be expected for  $\sigma=1522$ km s$^{-1}$. 
In other words in an isothermal potential
the mass distribution of the cluster does not correspond to the
isotropic one-dimensional velocity dispersion.
\rx ~may be an example of cluster which has not reached virial
equilibrium, its dynamical state may be in large part dominated by
infall or merging and consequently the velocity dispersion is not
representative of the virial temperature of the cluster. 
If there is any fraction of infalling galaxies which are bound to the
cluster but not yet virialized, they could inflate the velocity
dispersion. These galaxies in the cluster would be moving
on radial orbits. It might well be that  galaxies are infalling towards the
center of the cluster, and while some galaxies have already reached
(or crossed) the core region, some others are still moving along
radial orbits towards its center. This is somewhat supported by the
gradient in velocity dispersion which is seen between eleven galaxies 
to the NE and nine galaxies to the SW (a small effect but significant 
at 2.7$\sigma$). Numerical simulations  have shown that
the universe is composed of a web of filaments and voids on the scale
of hundreds of megaparsecs. The clusters of galaxies form at the
intersections of these filaments and grow with time
from quasi-spherical systems to the spherical objects that
we observe today. The initial formation of protoclusters
is often described as matter flowing along filaments (Bond et al. 1996)
with the X-ray flux maximum at the impact point of two colliding streams
of matter. In this scenario the hot gas and the galaxies would not be in
hydrostatic equilibrium and the transient velocity dispersion could
be higher than expected from a virialized system. We are probably
witnessing this process in this very distant X-ray selected cluster,
and \rx ~might be more properly called a protocluster.

\acknowledgments
I acknowledge the support and dedication of Chris Mullis and Pat Henry 
to the NEP project. This work would not have been possible without the
collaboration of John Huchra and several MPE scientists. Partial financial 
support comes from NSF grant AST95-00515 and from CNR-ASI grants. Finally,
I wish to thank Alain, Olivier, Vincent and the other LOC people for
organizing a great meeting in a very beautiful city.

\newpage
\vskip 1truecm

\begin{figure}
\centerline{**************see 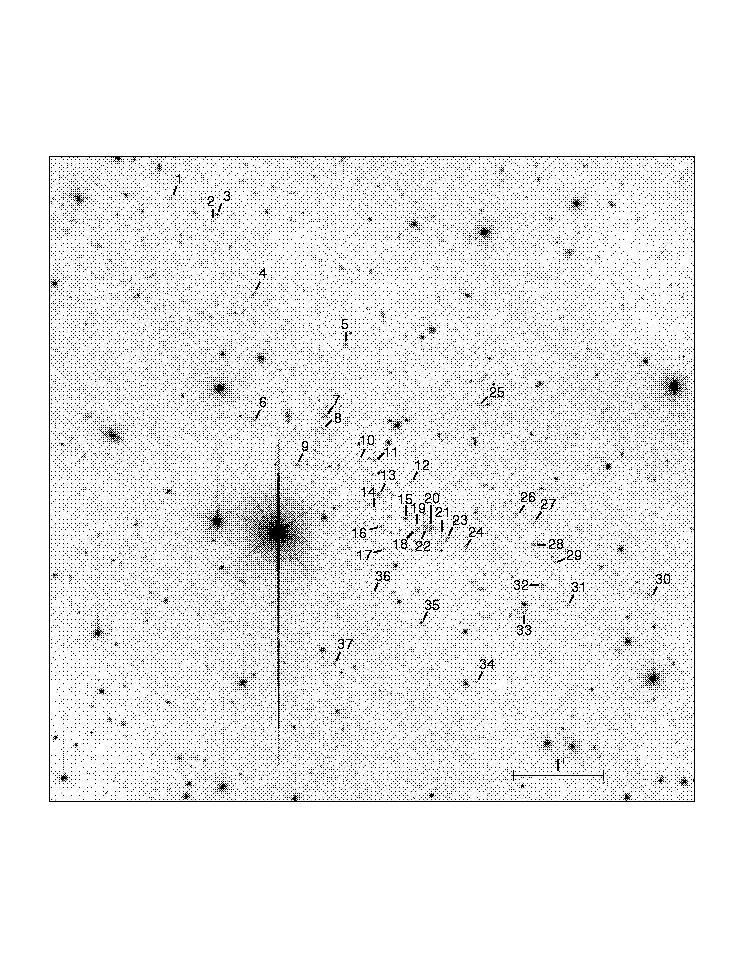**************}
\caption{The image is a 1024$\times$1024 subarray extracted from the
center of a 4500s exposure in the I-band taken by Luppino
and Metzger with the UH 8K$\times$8K CCD mosaic-camera on the CFHT
prime focus. Thirty-seven cluster members with Keck-II spectroscopy
are marked.  North is up and East to the left.}
\end{figure}

\begin{figure}
\centerline{**************see 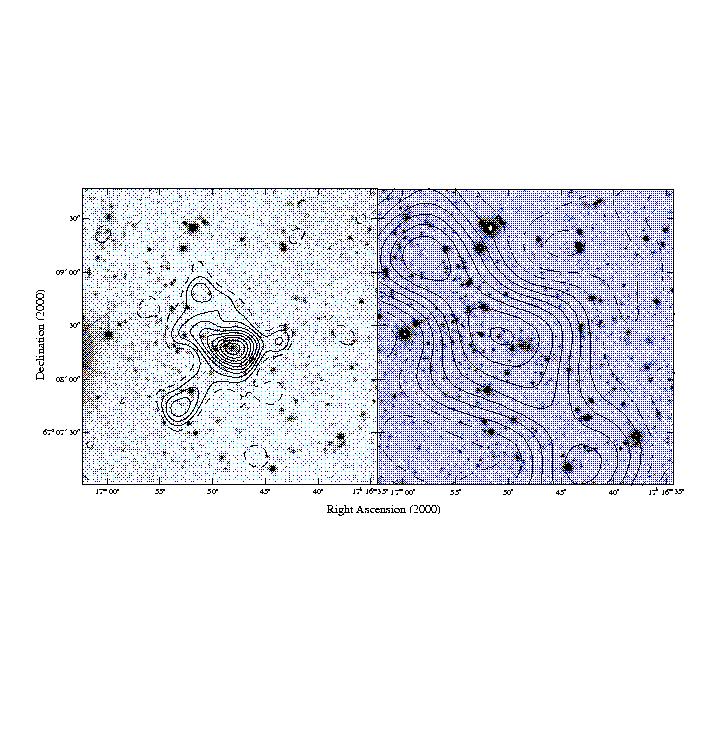**************}
\caption{Left panel shows the core of the cluster with contours of the 
HRI flux overlaid, while the right panel shows a  Keck-II R-band image 
of the same area with the weak lensing contours from Clowe et al. (1998)
overlaid. Either   panel covers an area of $2.75\times 2.75$
arcmin$^2$ ($1.3\times 1.3$ h$^{-2}_{50}$ Mpc$^2$ at the
redshift of the cluster).}
\end{figure}

\end{document}